%% file: thaller-psm-research.tex
\documentclass[10pt, conference]{IEEEtran}
\IEEEoverridecommandlockouts

\usepackage{setspace}
\usepackage[utf8x]{inputenc}
\usepackage[T1]{fontenc}
\usepackage[english]{babel}
\usepackage{microtype}
\usepackage{amsmath,amssymb,amsfonts, bm, amsthm}
\usepackage{algorithmic}
\usepackage{graphicx, tikz}
\usepackage{booktabs, threeparttable, multirow}
\usepackage{stfloats}
\usepackage[inline]{enumitem}
\usepackage{url}
\usepackage{xcolor}
\usepackage{caption, subcaption}
\usepackage{hyperref}
\usepackage{pifont}
\usepackage[framemethod=tikz]{mdframed}
\usepackage{siunitx}
\usepackage{balance}

\graphicspath{{images/}, {scripts/}}
\DeclareGraphicsExtensions{.pdf,.jpeg,.png}

\definecolor{dark-red}{rgb}{.6, .15, .15}
\definecolor{dark-blue}{rgb}{.15, .15, .55}
\definecolor{accent1}{HTML}{24B9FC}
\definecolor{accent2}{HTML}{9ECD67}
\definecolor{accent3}{HTML}{FFA929}
\definecolor{light-bg}{HTML}{B2B2B2}

\newcommand{\ra}[1]{\renewcommand{\arraystretch}{#1}}

\hypersetup{
	colorlinks,
  linkcolor={dark-red},
	citecolor={dark-blue},
  urlcolor={black}
}

\newcommand{\code}[1]{\texttt{#1}}

\newif\ifunblind
\unblindtrue

\begin{document}

\title{Probabilistic Software Modeling: \\A Data-driven Paradigm for Software Analysis}

\ifunblind
\author{%
	\IEEEauthorblockN{Hannes Thaller, Lukas Linsbauer, Alexander Egyed}
	\IEEEauthorblockA{Institute for Software Systems Engineering\\
		Johannes Kepler University Linz, Austria\\
		\{hannes.thaller, lukas.linsbauer, alexander.egyed\}@jku.at} \and
	\IEEEauthorblockN{Rudolf Ramler}
	\IEEEauthorblockA{Software Competence Center Hagenberg GmbH\\
		Austria\\
		rudolf.ramler@scch.at}
}
\fi

\maketitle

\input{vars}

\begin{abstract}
Software systems are complex, and behavioral comprehension with the increasing amount of AI components challenges traditional testing and maintenance strategies.
The lack of tools and methodologies for behavioral software comprehension leaves developers to testing and debugging that work in the boundaries of known scenarios.
We present Probabilistic Software Modeling (PSM), a data-driven modeling paradigm for predictive and generative methods in software engineering.
PSM analyzes a program and synthesizes a network of probabilistic models that can simulate and quantify the original program's behavior. 
The approach extracts the type, executable, and property structure of a program and copies its topology.
Each model is then optimized towards the observed runtime leading to a network that reflects the system's structure and behavior.
The resulting network allows for the full spectrum of statistical inferential analysis with which rich predictive and generative applications can be built.
Applications range from the visualization of states, inferential queries, test case generation, and anomaly detection up to the stochastic execution of the modeled system.
In this work, we present the modeling methodologies, an empirical study of the runtime behavior of software systems, and a comprehensive study on PSM modeled systems.
Results indicate that PSM is a solid foundation for structural and behavioral software comprehension applications.
\end{abstract}

\begin{IEEEkeywords}
probabilistic modeling, software modeling, static code analysis, dynamic code analysis, runtime monitoring, inference, simulation, deep learning, normalizing flows
\end{IEEEkeywords}

\IEEEpeerreviewmaketitle

\setstretch{.965}
\input{main}

\ifunblind
\section*{Acknowledgments}
The research reported in this paper has been supported by the Austrian Ministry for Transport, Innovation and Technology, the Federal Ministry of Science, Research and Economy, and the Province of Upper Austria in the frame of the COMET center SCCH.
\fi

\cleardoublepage

\setstretch{1}
\balance
\bibliographystyle{IEEEtran}
\bibliography{references}

\end{document}

%% file: vars.tex
\newcommand{\eTextScheme}{Doc2Vec \cite{Le2014}}
\newcommand{\eTextDim}{$10$}
\newcommand{\eConditionalScheme}{binary enconding}
\newcommand{\eConditionalDim}{$k$}
\newcommand{\eCouplings}{$6$}
\newcommand{\eLayers}{$1$}
\newcommand{\eUnits}{$256$}
\newcommand{\eVaePreprocessing}{standard score}
\newcommand{\eProjectCount}{4}
\newcommand{\eAspectj}{AspectJ 1.9.1}
\newcommand{\eJava}{Java 7}
\newcommand{\eRepetition}{$10$}
\newcommand{\pwAverage}[3]{Mdn\textsubscript{pw} = \num{#1}, IQR\textsubscript{pw} = \numrange{#2}{#3}}
\newcommand{\average}[3]{Mdn = \num{#1}, IQR = \numrange{#2}{#3}}
\newcommand{\pwMdnt}{Mdn\textsubscript{pw}}
\newcommand{\pwIQRt}{IQR\textsubscript{pw}}
\newcommand{\pwMdn}[1]{Mdn\textsubscript{pw} = \num{#1}}
\newcommand{\pwIQR}[2]{IQR\textsubscript{pw} = \numrange{#1}{#2}}
\newcommand{\IQR}[2]{IQR = \numrange{#1}{#2}}

%% file: main.tex
\section{Introduction}

Software complexity increases with every requirement, feature, revision, module, or software 2.0 (Artificial Intelligence (AI)) component that is integrated.
Complexity related challenges in traditional software engineering have many tools and methodologies that mitigate and alleviate issues (e.g., requirements engineering, version control systems, unit testing).
However, tight integration of AI components in programs is still in its infancy and so are the methodologies and tools that allow combined analysis, development, testing, integration, and maintenance.

We present \emph{Probabilistic Software Modeling (PSM)}, a data-driven modeling paradigm for predictive and generative methods in software engineering.
PSM is an analysis methodology for traditional software (e.g., Java \cite{Arnold2000}) that builds a \emph{Probabilistic Model (PM)} of a program.
The PM allows developers to reason about their program's semantics on the same level of abstraction as their source code (e.g., methods, fields, or classes) without changing the development process or programming language.
This enables the advantages of probabilistic modeling and causal reasoning for traditional software development that are fundamental in other domains (such as medical biology, material simulation, economics, meteorology).
PSM enables applications such as test-case generation, semantic clone detection, or anomaly detection seamlessly for both, traditional software as well as AI components and their randomness.
Our experiments indicate that PMs can model programs and allow for causal reasoning and consistent data generation that these applications are built on.

PSM has four main aspects: \emph{Code (Structure), Runtime (Behavior), Modeling, and Inference}.
First, PSM extracts a program's \emph{structure} via static code analysis (\emph{Code}).
The abstraction level is properties, executables, and types (e.g., fields, methods, and classes in Java) but ignores statements, allowing PSM to scale.
Second, it inspects the program's \emph{behavior} by observing its runtime (\emph{Runtime}).
This includes property accesses and executable invocations.
Then, PSM combines this static structure and dynamic behavior into a probabilistic model (\emph{Modeling}).
This step also represents the main contribution of this work.
Finally, predictive or generative applications (e.g., a test-case generator or anomaly detector) leverage the models via statistical inference (\emph{Inference}).

The prototype used for the evaluation is called \emph{Gradient}\footnote{\ifunblind\url{https://github.com/jku-isse/gradient}\fi} and is openly available.

First, Section~\ref{sec:related work} views our contribution from the perspective of existing related domains.
Section~\ref{sec:example} introduces an illustrative example we use throughout this paper.
In Section~\ref{sec:applications} we motivate our contribution by providing an outlook on possible applications and research opportunities that PSM enables.
Then we briefly discuss the nomenclature and background needed to understand PSM (Section~\ref{sec:background}).
Section~\ref{sec: approach} presents the main contribution containing the general usage pragmatism and construction methodologies for PSM models on a conceptual level.
A comprehensive evaluation of whether software can be transformed into statistical models is given in Section~\ref{sec:study} and discussed in Section~\ref{sec:discussion}.
Section~\ref{sec:conclusion} concludes the paper.

\section{Related Work}\label{sec:related work}
To position PSM it is useful to distinguish between \emph{programming paradigms} and \emph{software analysis methods}.
A programming paradigm is a collection of programming languages that share common traits (e.g., object-oriented, logical, or functional programming).
Analysis methods extract information from programs (e.g., design pattern detection, clone detection).
PSM is an \emph{analysis} method that analyzes a program given in an object-oriented programming language and \emph{synthesizes} a probabilistic model from it.

\emph{Probabilistic programming} is a programming paradigm in which probabilistic models are specified.
Developers describe probabilistic programs in a domain-specific language (e.g., BUGS~\cite{Lunn2009}) or via a library in a host language (e.g., Pyro~\cite{Bingham2019}, PyMC~\cite{Salvatier2016}, Edward~\cite{Tran2017b}).
In contrast, PSM analyzes a program written in a traditional programming language and translates it into a probabilistic program.
This difference also holds for modeling concepts like \emph{Bayesian Networks} \cite{Koller2009} or \emph{Object-Oriented Bayesian Networks} \cite{Koller1997, Musella2015} that can be implemented via a probabilistic programming language.

\emph{Formal methods} are a programming paradigm that leverages logic as a programming language (e.g., TLA+~\cite{Lamport2002} or Alloy~\cite{Jackson2002}).
\emph{Stochastic model checking}~\cite{Kwiatkowska2007} introduces uncertainty in the rigid formalism to model, e.g., natural phenomenons.
Developers specify the behavior and provide the state transition probabilities in a special-purpose language (e.g., PRISM~\cite{Kwiatkowska2011}, PAT~\cite{Liu2011}, CADP~\cite{Garavel2013}).
Again, PSM analyzes a program and synthesizes a PM allowing developers to work with the programming language of their choice.

\emph{Symbolic execution}~\cite{King1976} is an analysis method that executes a program with symbols rather than concrete values (e.g., JPF-SE~\cite{Anand2007}, KLEE~\cite{Cadar2008}, Pex~\cite{Tillmann2008}).
It can be used to determine which input values cause specific branching points (if-else branches) in a program.
\emph{Probabilistic symbolic execution}~\cite{Geldenhuys2012} is an extension that quantifies the execution, e.g., branching points, in terms of probabilities.
This is useful for applications that quantify program changes~\cite{Filieri2015} or performance~\cite{Chen2016}.
Probabilistic symbolic execution operates on the statement level while PSM abstracts statements capturing, e.g., inputs and outputs of methods.
This abstraction makes PSM computationally scalable while symbolic execution suffers from state explosions.
Furthermore, this abstraction shifts the analysis focus to the program semantics compared to the statement semantics (e.g., what happens between methods vs. what happens at the if statement).

\emph{Probabilistic debugging}~\cite{Xu2018, Andrzejewski2007} is an analysis method that supports developers in debugging sessions.
The debugger assigns probabilities to each statement and updates them according to the most likely erroneous statement.
Again, in contrast to PSM, they operate on statement level.
Another difference is given in the methodologies life cycle.
Debugging has an operational life cycle only valid until the bug is found.
PSM and the resulting models are intended to be persisted along with the matching source code revision.
This allows, e.g., method-level error localization, by comparing multiple revisions of the same model.

\emph{Invariant detectors} \cite{Hangal2002, Ernst2001, Le2018, Zuo2014, Gore2011, Lo2012} learn assertions and add them to the source code.
This helps to pinpoint erroneous regions in the source code.
Invariant detectors learn rules of value boundaries of statements (i.e., pre- and post-conditions), not the actual distribution.
However, this distribution allows PSM to generate new data enabling causal reasoning across multiple code elements.

\section{Illustrative Example}\label{sec:example}
\begin{figure*}
	\centering
	\begin{subfigure}[t]{0.37\textwidth}
		\includegraphics[width=\textwidth]{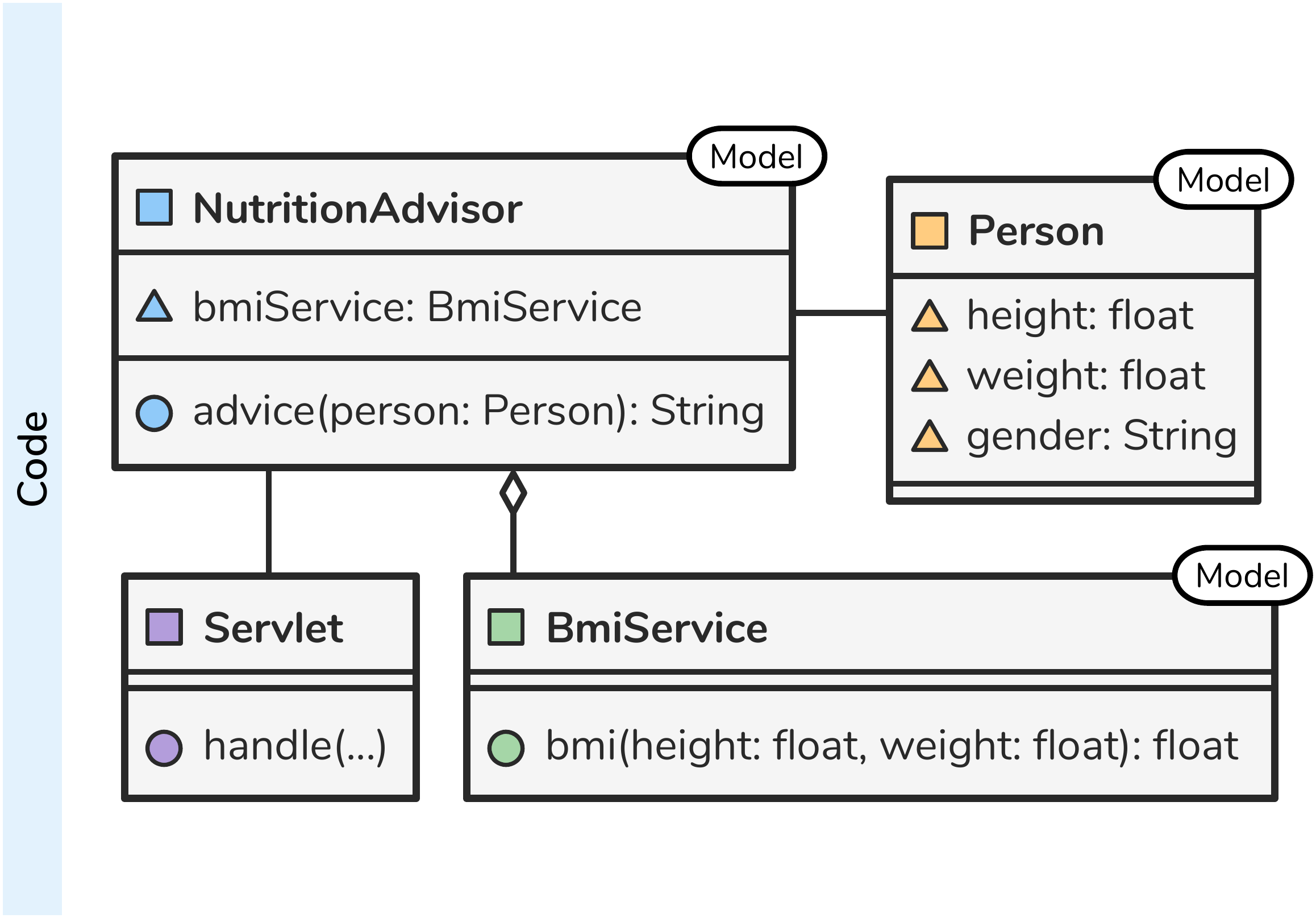}
		\caption{ 
			\textbf{[Code]} The static structure of the Nutrition Advisor, consisting of three core classes and a context class (e.g., a web-interface) calling the program.
		}
		\label{fig: nutrition advisor structural}
	\end{subfigure} \hfill
	\begin{subfigure}[t]{0.62\textwidth}
		\includegraphics[width=\textwidth]{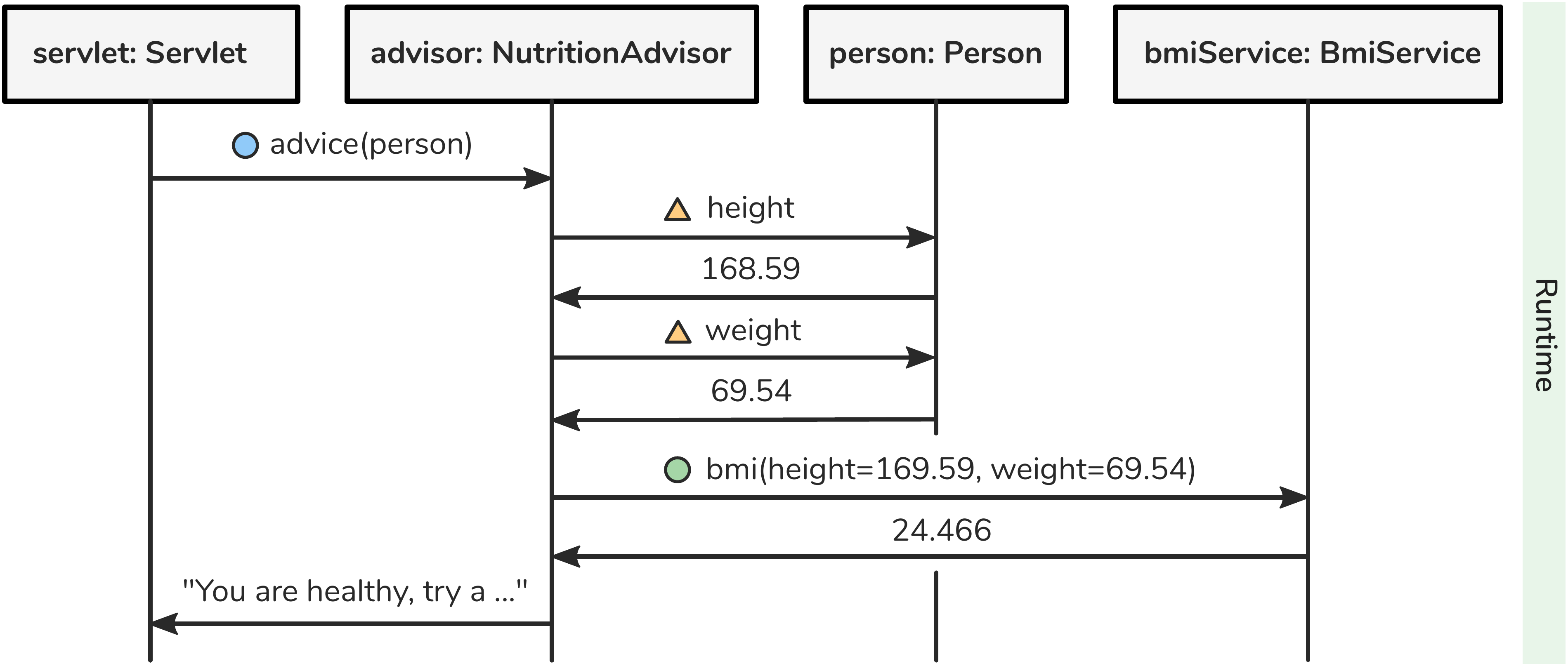}
		\caption{
			\textbf{[Runtime]} The dynamic behavior of the Nutrition Advisor, visualized by one execution trace.
			The \code{NutritionAdvisor} handles \code{advice} requests in which \code{Person} objects are received and a textual advice is returned.
		}
		\label{fig: nutrition advisor behavioral}
	\end{subfigure}
	\caption{
		The Nutrition Advisor receives a person with its anthropometric measurements and computes a textual advice regarding the person's diet.
		For simplicity some properties and executables are omitted.
	}
	\label{fig:nutrition advisor}
\end{figure*}

Consider as our running example the \emph{Nutrition Advisor} that takes a person's anthropometric measurements (height and weight) and returns a textual advice based on the \emph{Body Mass Index (BMI)}.
Figure~\ref{fig: nutrition advisor structural} shows the class diagram of the Nutrition Advisor, consisting of three core classes and the \code{Servlet} class.
Classes considered by PSM are annotated with \emph{Model} (e.g., \code{Person}).
Figure~\ref{fig: nutrition advisor behavioral} depicts a sequence diagram of one program trace with concrete values.
The \code{Servlet} receives properties (e.g., height, weight, or gender) with which it instantiates a Person object (not shown).
\code{NutritionAdvisor.advice($\cdot$)} takes this \code{Person} object, extracts the \code{height} (\num{168.59}) and \code{weight} (\num{69.54}) and computes the person's BMI (\num{24.466}) via \code{BmiService.bmi($\cdot$)}.
The result is a textual advice based on the BMI ("You are healthy, try a ...").
Note that, for the sake of simplicity, Figure~\ref{fig: nutrition advisor structural} only shows a subset of the code elements from the real Nutrition Advisor (e.g., \code{Person.name} or \code{Person.age} are omitted).
Given a program such as the Nutrition Advisor, PSM can be used to build a network of probabilistic models with the same structure and behavior.

\section{Motivating Applications}\label{sec:applications}
PSM is a generic framework that enables a wide range of predictive and generative applications.
This section lists a selection of possible applications.

\subsection{Predictive Applications}
Predictive applications seek to quantify, visualize, infer and predict the behavior and quality of a system.

\textbf{Visualization and Comprehension}~\cite{Jayaraman2017, Brown1985, Mukherjea1994} applications help to understand software and its behavior.
This includes the visualization of code elements and non-functional attributes such as performance.
The PMs are the source of the visualization showing the global but also contextual behavior across code elements.
For example, Figure~\ref{fig:nutrition advisor prob behavioral} visualizes the \code{height}-property in which typical and less typical values can be seen in a blink.
$P(Height \mid Gender = Female)$ visualizes a context-aware behavior how gender affects the height.

\textbf{Semantic Clone-Detection}~\cite{Gabel2008, Kim2011} applications detect syntactically different but semantically equivalent code fragments, e.g., the iterative and recursive version of an algorithm.
Traditionally, clone detection compares source code fragments focusing on exact or slightly adapted clones.
However, semantic equality is beyond purely static properties of source code.
PSM can detect method level clones by comparing their models.
The comparison can be realized, for example, via statistical tests on sampled data~\cite{Mann1947, Kruskal1952, Massey1951} (simple automated decision), via visualization techniques such as Q-Q plots~\cite{Wilk1968} (comprehensive manual decision), or a combination these.

\textbf{Anomaly Detection}~\cite{Hangal2002, Aniello2016, Kotu2019, Chandola2009} applications measure the divergence between a persisted PSM model and a newly collected observation.
These applications can be deployed into a live system, in which components are monitored and checked against their models.
A threshold checks for unlikely runtime observations $x$ (i.e., $p(Weight=weight_{new}) < .1$) triggering additional actions in cause of a failure.
$x$ and its effects on other elements can then be investigated with, e.g., visualization and comprehension techniques, for further decision-making processes.

\subsection{Generative Applications}
Generative applications leverage observations drawn from the models, e.g., executable inputs or property values.

\textbf{Test-Case Generation}~\cite{Cseppento2017, Fraser2012} applications draw observations from executable and property models to generate test data.
PSM can generate \emph{scoped} test data with a specific likelihood or for a specific system scenario (system state).
For instance, likelihood-scoped data can be used to generate different test suites such as \emph{typical, rare,} or \emph{unseen} by sampling $x < P(Person) = P(Height, Weight) < y$ where $x$ and $y$ are predefined boundaries of the likelihood.
This helps to strengthen test suites with meaningful, automatically generated tests based on real (un)likely behavior.

\textbf{Simulation} applications sample execution traces from the network of models in a structured fashion to reproduce the running system.
This probabilistically executes the original program without actually running it.
Simulations can bridge boundaries between hardware and software interfaces, reducing the number of hardware dependencies during development.

\section{Background}\label{sec:background}
\begin{figure*}
	\centering
	\begin{subfigure}[t]{0.55\textwidth}
		\includegraphics[width=\textwidth]{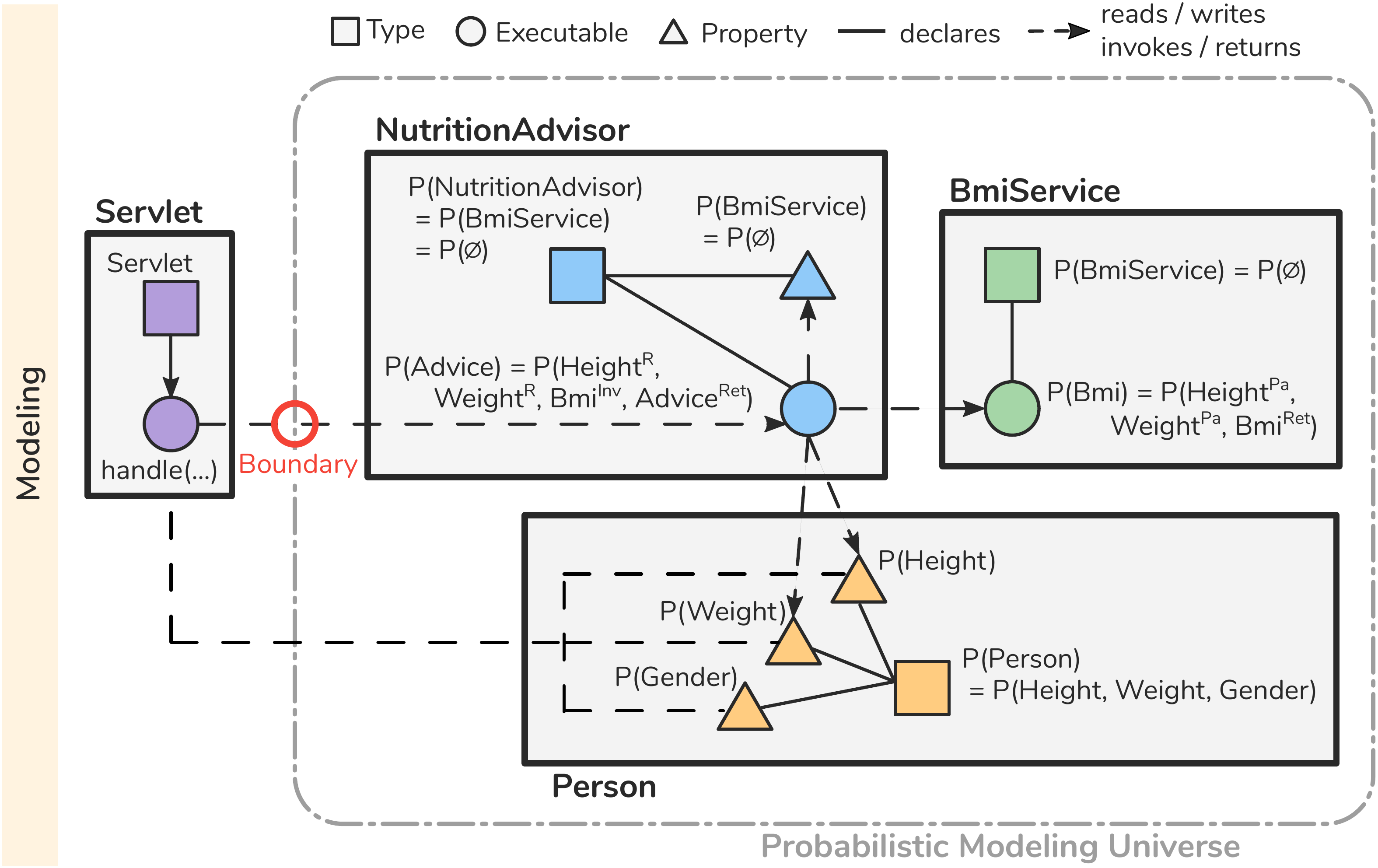}
		\caption{ 
			\textbf{[Modeling]} The Probabilistic Model Network of the Nutrition Advisor (simplified).
			Elements within the Probabilistic Modeling Universe are modeled according to their probabilistic expressions.
			Triangles are properties, circles are executables, and rectangles are types.
			The superscripts represent property reads $R$, and executable invocations $Inv$, parameters $Pa$, and return values $Ret$.
		}
		\label{fig:nutrition advisor prob structural}
	\end{subfigure} \hfill
	\begin{subfigure}[t]{0.43\textwidth}
		\includegraphics[width=\columnwidth]{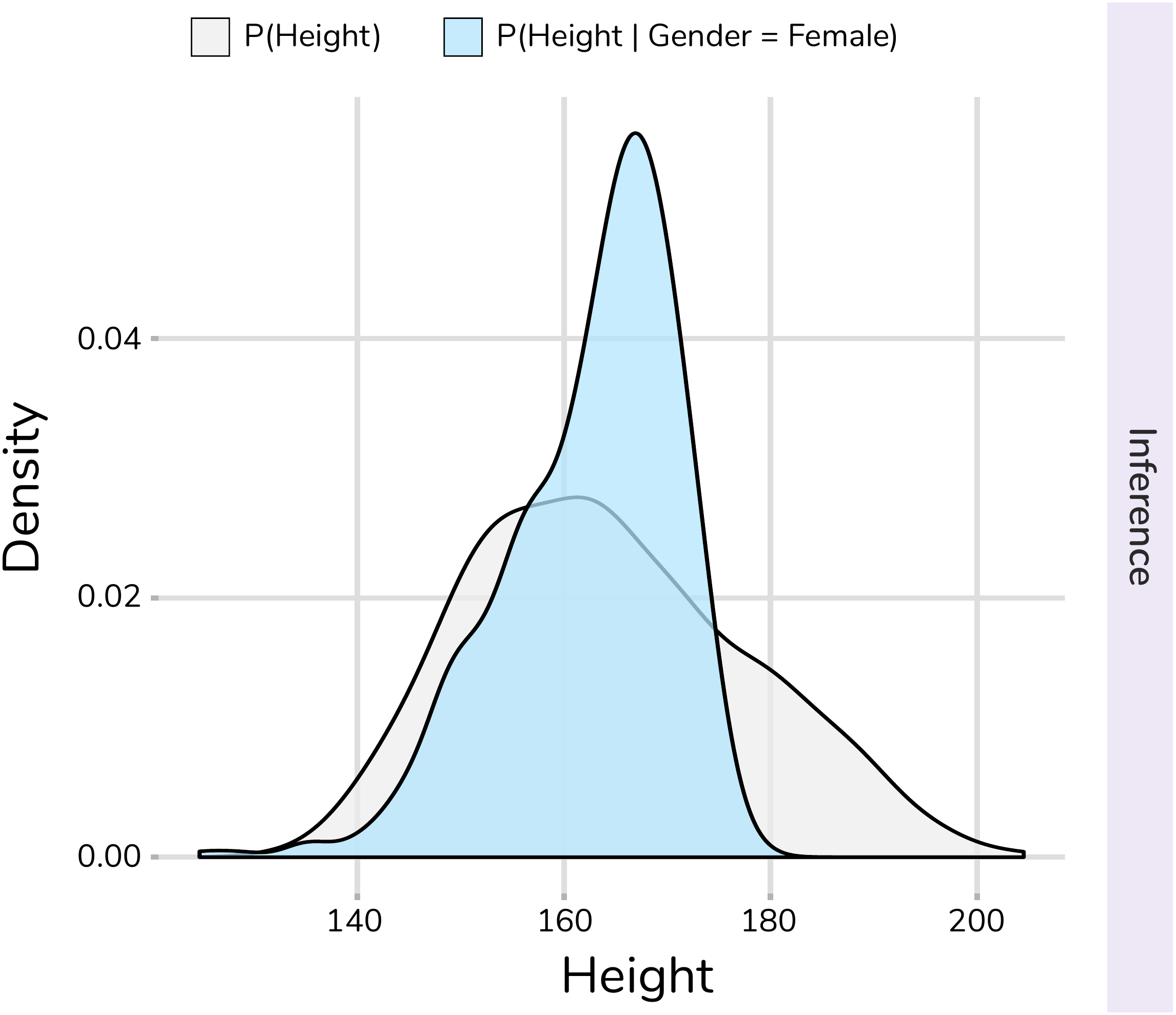}
		\caption{ 
			\textbf{[Modeling]} The distribution of the \code{Person.weight} properties.
			The histogram are the runtime observations that were sampled from the \emph{True Distribution} (usually unknown).
			The \emph{Fitted Distribution} is the model approximation based on the data.
			}
		\label{fig:nutrition advisor prob behavioral}
	\end{subfigure}
	\caption{The Nutrition Advisor system as \emph{Probabilistic Model Network} (left) and the model of the \code{Person.weight} node (right).}
	\label{fig: nutrition advisor prob}
\end{figure*}

PSM combines two major domains: Software Engineering (SE) and Machine Learning (ML).
Naturally, some terms can be misinterpreted depending on the readers background.
The following terminology was chosen as the best common ground and might be untypical in the respective domain.

\subsection{Code}
Types, properties, and executables are object-oriented terms (e.g., classes, fields, and methods in Java~\cite{Arnold2000}, see Figure~\ref{fig: nutrition advisor structural}).
In the context of PSM, these are referred to as \emph{code elements}.
These code elements can be organized in an \emph{Abstract Semantics Graph (ASG)}, which is a high-level version of an abstract syntax tree (AST).
An ASG contains no lexical nodes but has additional semantic relationships (e.g., typing information of expressions).
Also, in the context of PSM, we define that each code element has a \emph{symbol}.
A symbol is a numerical identifier, e.g., $Symbol(Person.weight) = 0$.

\subsection{Runtime}
Runtime monitoring (or dynamic code analysis) \cite{Ball1999} is the process of observing a running program.
The program is executed by a \emph{trigger} (parameters and environment) which is the context of the monitoring session.
A running program spawns \emph{event streams} which are sequences of \emph{monitoring events} (e.g., Figure~\ref{fig: nutrition advisor behavioral}).
These events contain information such as properties that were changed or executables that were invoked.
Also, the stream shows which parts of the underlying source code are active with the given trigger.
\emph{Tracing} tracks every possible event at runtime, whereas \emph{sampling} records events according to a specific rate.

\subsection{Modeling}
A \emph{probabilistic model} uses the theory of probability to model a complex system (e.g., Nutrition Advisor).
A \emph{random variable} $X_i \in \mathcal{X}$ (e.g., $Weight$) captures an aspect of the system's event space.
The value range of random variables is given by $Val(X_i)$ (e.g., $Val(Weight) = \{i \mid i \in \mathbb{R}\}$).
A \emph{probability distribution} $P$ is a mapping from events in the system to real values (e.g., Figure~\ref{fig:nutrition advisor prob behavioral} histogram elements map to a point on the \emph{Fitted Distribution} line).
These values are between $0$ and $1$ and all values sum up to 1.
The \emph{marginal distribution} $P(X_i)$ describes the probability distribution of the random variable $X_i$ (e.g., $P(Weight)$).
The \emph{joint distribution} $P(X_1, \ldots, X_n)$  represents the probability distribution that can be described with all of the variables (e.g., $P(Weight, Height)$).
A \emph{conditional distribution} $P(X \mid Y)$ describes the probability distribution of $X$ given that some additional information of the random variable $Y$ was observed (e.g., $P(Weight \mid Height = 193cm)$).
$Y$ is called the \emph{conditional} and scopes the distribution of $X$.
More background information is given, e.g., by Koller and Friedman~\cite{Koller2009}, Murphy~\cite{Murphy2012}, or Bishop~\cite{Bishop2006}.

PSM is mostly interested in the conditional distribution of a code element given its invoking context, e.g., a property access $P(\text{Weight} \mid C)$ with its context $C$ (e.g., advice method).
A probabilistic expression such as $P(Weight \mid C)$ is equivalent to pseudocode in SE.
They describe a process (e.g., a sorting algorithm) that can be parameterized with a concrete implementation and technology (e.g., functional implementation given in Haskell \cite{PeytonJones2003} or object-oriented implementation in Java \cite{Arnold2000}).
Similarly, $P(\text{Weight} \mid C)$ can be parameterized via a stochastic model representing its quantity, and in hindsight its process.

This work presents the modeling strategies (see Section~\ref{sec: approach}) in the form of probabilistic expressions that our prototype parameterizes via Real Non-Volume Preserving Transformations (NVPs) \cite{Dinh2016}.
NVPs are density estimators that allow efficient and exact inference, sampling, and likelihood estimation of data points.
NVPs learn an invertible and pure bijective function $f: X \mapsto Z $ (with $g = f^{-1}$) that map the original input variable  $x \in X$ to simpler latent variables $z \in Z$.
The latent variables are often isotropic unit norm Gaussian $N(0, \bm{1})$ that are well understood in terms of sampling and likelihood evaluation.
An NVP is a combination of multiple small neural networks, called coupling layers, that are combined by simple scale and translation transformations.
\emph{Conditional NVPs} are an extension that estimate $P(X \mid C)$.

\subsection{Inference}
Every PSM application in Section \ref{sec:applications} is build upon inference.
It is the combination of \emph{sampling}, \emph{conditioning}, and \emph{likelihood evaluation}.

Each node in a PSM network is an NVP.
Sampling with NVPs is done by sampling from the Gaussian latent-space $z \sim N(0, \bm{1})$ and applying the NVP in inverse $x = g(z)$.
NVP can be conditioned statically and dynamically.
Static conditioning is achieved by adding additional features to the network during training.
Dynamic conditioning finds latent-space configurations that match the condition by, e.g., variational inference \cite{Blei2017, Rezende2015}.
Finally, likelihood evaluation is achieved by evaluating the likelihood under the Gaussian latent-space times the NVPs Jacobian
\begin{equation}\label{eq: nvp logp}
	p_X(x) = p_Z\left(f\left(x\right)\right)   \left\vert \det\left(\dfrac{\partial f(x)}{\partial x^T}\right)   \right\vert.
\end{equation}
More details are given by Dinh et al. \cite{Dinh2016}.

\section{Approach}\label{sec: approach}
\begin{figure*}[t]
	\centering
	\includegraphics[width=.9\linewidth]{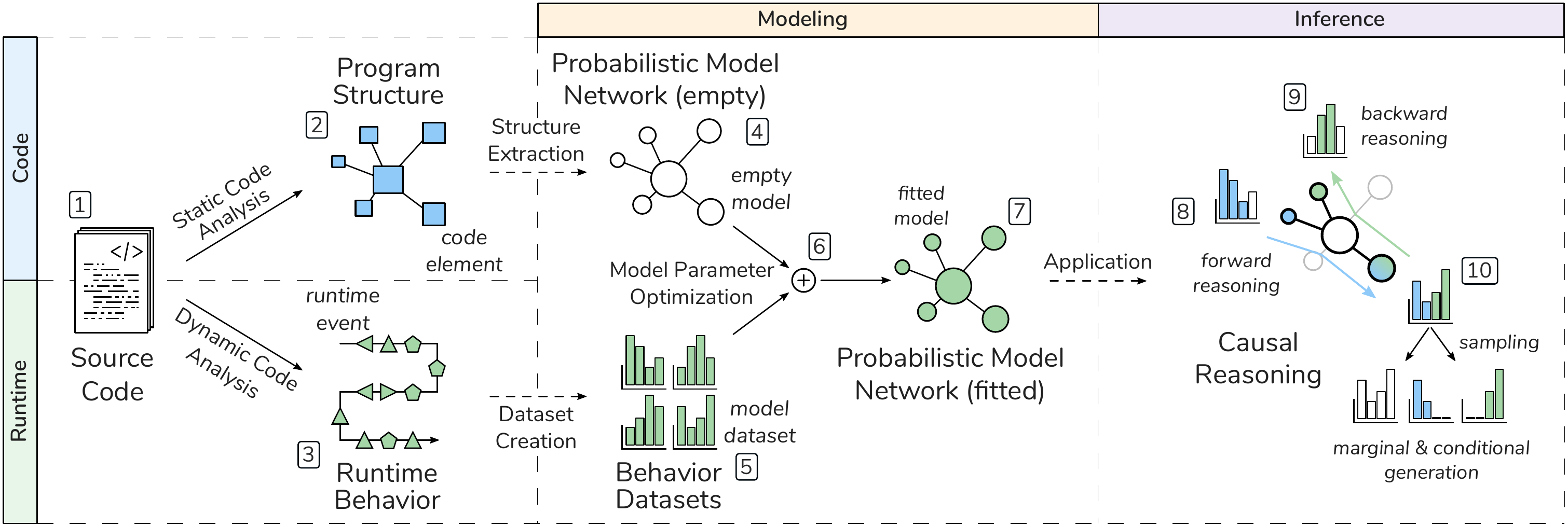}
	\caption{
		\emph{Source Code} (1) has a \emph{Program Structure} (2) and a \emph{Runtime Behavior} (3) that is extracted via \emph{Static} and \emph{Dynamic Code Analysis}.
		These result in a \emph{Probabilistic Model Network (empty)} (4) and \emph{Behavior Datasets} (5) that are combined by \emph{Optimization} (6) into the final \emph{Probabilistic Model Network (fitted)} (7).
		}
	\label{fig:overview}
\end{figure*}

PSM is a four-fold approach illustrated in Figure~\ref{fig:overview} in which:
\begin{enumerate}
	\item \textbf{[Code]} static code information is extracted and analyzed;
	\item \textbf{[Runtime]} runtime behavior is collected and transformed;
	\item \textbf{[Modeling]} probabilistic models are built by combining code and runtime data;
	\item \textbf{[Inference]} applications are build by leveraging causal reasoning and data generation. 
\end{enumerate}
The main contributions of this work are concepts and realizations in the \emph{Modeling} aspect.

\subsection{Code}\label{sec: approach code}
The input is the \emph{Source Code} (1) of a program (e.g., of the Nutrition Advisor).
Then, \emph{Static Code Analysis} extracts the \emph{Program Structure} (2) in the form of an ASG.
The class diagram in Figure~\ref{fig: nutrition advisor structural} may act as an abstract substitute of the structure in this example.
Elements that are to be modeled are annotated with the label \emph{Model}.
In that regard, PSM is selective of the code elements considered for static and dynamic code analysis.
The selection depends on the application context (see Section~\ref{sec:applications}), or the developer's interest.
The set of all code elements PSM considers is called the \emph{Modeling Universe}.

\subsection{Runtime}\label{sec: approach Runtime}
\emph{Dynamic Code Analysis} extracts the \emph{Runtime Behavior} (3) by executing the program with a trigger and monitoring the internal events.
This results in an event stream similar to the sequence diagram in Figure~\ref{fig: nutrition advisor behavioral}.
Events are property accesses and executable invocations of code elements in the modeling universe.
Depending on the application context (see Section~\ref{sec:applications}), execution triggers can be, e.g., tests (weak), or the runtime of a deployed system (strong).
For example, \emph{Visualization and Comprehension} demands a trigger as close as possible to the real environment (manual understanding).
In contrast, \emph{Semantic Clone-Detection} makes differential comparisons between models where synthetic data suffices (automatic comparisons).

\subsection{Modeling}\label{sec: approach modeling}

PSM extracts from the \emph{Program Structure} the code element topology and builds the \emph{Probabilistic Model Network (empty)} (see Figure \ref{fig:overview}, step 4).
From a software engineering perspective, this process is comparable to traversing the ASG and attaching an empty (unfitted) PM to the every node.
An example network is demonstrated in Figure~\ref{fig:nutrition advisor prob structural} where each node is a PM.
The actual construction rules (probabilistic expressions) to build such a PSM network are given below (Section \ref{sec: construction rules}).
The \emph{Dataset Creation} tallies and pre-processes the event stream into \emph{Behavior Datasets} (5) for each code element.
The \emph{Model Parameter Optimization} (6) fits each PM, i.e., node in the \emph{Probabilistic Model Network}, to the \emph{Behavior Dataset} of the associated code element.
This results the \emph{Probabilistic Model Network (fitted)} (7) with the same topology found in the \emph{Program Structure}, optimized towards the observed \emph{Runtime Behavior}.

\subsubsection{Construction Rules}\label{sec: construction rules}
The \emph{construction rules} define how each node in the \emph{Probabilistic Model Network} (4), i.e., a given code element, is transformed into a probabilistic expression.
This expression is a description of the model (random) variables and its approximating quantity (e.g., see Figure \ref{fig:nutrition advisor prob structural}).
Hence, building the PM network equals
\begin{enumerate*}
	\item a traversal in the program's ASG;
	\item an application of the construction rules creating a probabilistic expression (per node); and
	\item the parameterization of the expressions with a concrete model (e.g., VAEs).
\end{enumerate*}

The \textbf{\emph{property} construction rule} defines a property model by the property value itself, conditioned on the symbol of the accessing executable (conditional).
\begin{equation}
	P(\text{Property} \mid C) = P(R, W \mid C) \label{eq: property factor}
\end{equation}
$R$ and $W$ are the read and write accesses to the property.
For example, the \code{Person.weight} model is defined by $P(\text{Weight} \mid C)$.
The value range of the property depends on the property itself, whereas the range of the conditional is all (executable) symbols that exist in the project $\mathit{Val}(C) = \mathit{Symbols}(\mathit{Project})$.
This includes executable symbols that live outside the PSM Universe.
The conditional allows PSM to differentiate between call sites.
This allows each call site to have a different distribution.
For example, \code{NutritionAdvisorAdolesence} and \code{NutritionAdvisorAdult} use the \code{BmiService} leading to two slightly shifted weight distributions in the same model.

The \textbf{\emph{executable} construction rule} defines an executable model by a joint distribution of the inputs and outputs, conditioned on the symbol of the invoking executable.
\begin{align}
P(\text{Executable} \mid C) &= P(\mathcal{I} , \mathcal{O} \mid C) \\
&= P(\underbrace{\bm{Pa}, \bm{Inv}, \bm{R}}_{\mathcal{I}}, \underbrace{\bm{W}, \bm{Ret}}_{\mathcal{O}} \mid C) \label{eq: executable factor}
\end{align}
$\bm{Pa}$ are parameters, $\bm{Inv}$ are (executable) invocations, $\bm{R}$ are property reads, $\bm{W}$ are property writes, $\bm{Ret}$ are the return values, and $C$ are all (executable) symbols that exist in the project $\mathit{Val}(C) = \mathit{Symbols}(\mathit{Project})$.
An example would be the \code{bmi}-method with $P(Bmi) = P(Height^{Pa}, Weight^{Pa}, Bmi^{Ret} \mid C)$ where $Val(C) = \{Symbol(NutritionAdvisor.advice)\}$  (see Figure \ref{fig:nutrition advisor prob structural}).

The \textbf{\emph{type} construction rule} defines a type model by the joint distribution of properties the type declares, conditioned on the symbol of the accessing executable.
\begin{equation}
	P(Type) = P(\bm{Property}^{Type} \mid C) \label{eq: type factor}
\end{equation}
For example, a \code{Person} object is defined by $P(Height, Weight)$.
The type distribution is empty in the case where no properties exist as Figure~\ref{fig:nutrition advisor prob structural} shows for the \code{bmiService} property in \code{NutritionAdvisor}.
Sampling from a type distribution instantiates a new object of a given type by assigning the sampled values to the properties.

\subsubsection{Technical Modeling Considerations}\label{sec:realization}
PSM estimates the density of the values that code elements emit during runtime in the form of generative models.
It searches for a model from which new samples can be drawn, and that compresses the original monitoring data into a fixed set of parameters.
This goal stipulates a set of requirements with which the network nodes can be parameterized.
The model should be a \emph{scalable, parametric, decidable, generative, (conditional) density estimator}.
\begin{itemize}
	\item \emph{Scalable} such that it can handle the enormous amounts of data running systems produce.
	\item \emph{Parametric} such that it has a fixed set of parameters that can be stored and shared.
	\item \emph{Decidable} such that the parameter optimization has a clear convergence criterion.
	\item \emph{Generative} such that it allows for efficient sampling of the approximated distribution.
	\item A \emph{(Conditional) Density estimator} that is capable of approximating arbitrary data.
\end{itemize}
Besides, the learning process should be as robust as possible to reduce human intervention.
Each requirement is tied to functional (generative, density estimator) or non-functional (scalable, parametric, decidable) requirements of PSM.
One class of models that fit many of these requirements are likelihood-based deep generative networks like \emph{Variational Auto-Encoders (VAEs)}~\cite{Kingma2013, Doersch2016, Sohn2015} or  flow-based methods like the \emph{Real Non-Volumetric Preserving Transformation}~\cite{Dinh2016} and derivatives \cite{Grathwohl2018, Germain2015a, Papamakarios2019} .

Another technical consideration is that Equations \ref{eq: property factor}, \ref{eq: executable factor}, and \ref{eq: type factor} can be factorized in each other.
That is, a real implementation does not need a model for each property, executable, and type but may combine them into one model.
The prototype in this work uses exclusively executable models (see Section \ref{sec:study}).

\subsection{Inference}
Inference is the fundament of all applications motivated in Section \ref{sec:applications} and illustrated in Figure \ref{fig:overview}.
The three tightly connected main aspects of inference are \emph{sampling} (generation), \emph{conditioning} (information propagation), \emph{likelihood evaluation} (criticism).
\emph{Sampling} draws observations from one (local) or multiple (global) nodes (NVPs) in the PSM network.
This enables the probabilistic execution of e.g., an executable or a subsystem.
\emph{Conditioning} sets the models into a specific state.
For example, Figure~\ref{fig:nutrition advisor prob behavioral} illustrates the height property in its unconditioned and conditioned state.
Local conditioning sets one node into a state.
Global conditioning propagates a state across multiple nodes.
\emph{Likelihood Evaluation} quantifies samples in terms of their likelihood under a given node (i.e., a model).

Figure~\ref{fig:overview} illustrates the combination of these aspects and combines them into causal forward (8) and backward (9) reasoning.
Forward reasoning (8) (e.g., \code{Person.height} to \code{BmiService.bmi}) samples a conditional distribution and propagates it through the network to set downstream nodes into a conditioned state.
Backward reasoning (9) starts at a conditioned downstream node and searches for the most likely cause.
At every step it is possible to draw conditional or unconditional samples.
The directional aspect (forward and backward) is based on the source codes dependency graph.
PSM networks, however, are undirectional (a network of joint-distributions).

\section{Study}\label{sec:study}
The core hypothesis of PSM is that programs can be transformed into a probabilistic model.
This study (i.e., the prototype, research questions, analyses, and discussions) focuses on evaluating the core PSM methodologies presented in Section~\ref{sec: approach}.
Specifically, the study answers the following questions, providing evidence for the core hypothesis:
\begin{enumerate}[label=\footnotesize{RQ}\arabic*]
\item \label{rq:code} \textbf{[Code]} Are projects exposing enough code elements that are eligible for PSM?
\item \label{rq:monitoring} \textbf{[Runtime]} Are code elements creating enough runtime data with which the model parameters can be optimized?
\item \label{rq:modeling} \textbf{[Modeling]} Are probabilistic models capable of capturing the runtime data of eligible code elements?
\item \label{rq:inferenece}\textbf{[Inference]} Is the network of probabilistic models capable of solving inferential tasks?
\end{enumerate}

\ref{rq:code} addresses the precondition whether projects expose enough data (i.e., number or text) code elements that can be modeled.
\ref{rq:monitoring} addresses the precondition whether these (data) code elements create a sufficient amount of runtime data that can be modeled.
\ref{rq:modeling} addresses the central question whether the behavior of a program in the form of its runtime data can be approximated via the concrete models.
Finally, \ref{rq:inferenece} evaluates the usefulness of the approach and whether PSM is a sound basis for the applications presented in Section~\ref{sec:applications}.
The four questions are scoped by structured programs that can be executed and support runtime monitoring.
The empirical evidence in this work is essential for any future endeavor related to statistical modeling of software.
The evaluation of concrete applications of PSM described in Section~\ref{sec:applications} are beyond the scope of this study.

\subsection{Setup}\label{sec: study setup}
\input{tab_model_parameters.tex}

We implemented a prototype called Gradient\footnotemark{} that reflects the process and data flow presented in Figure~\ref{fig:overview}.
\footnotetext{\ifunblind\url{https://github.com/jku-isse/gradient-benchmark}\fi}
\begin{enumerate}
	\item The input \emph{Source Code} were open source subject systems written in Java (see next Section~\ref{sec: subject systems}).
	\item The \emph{Program Structure} was extracted using Spoon~\cite{Pawlak2015}.
	\item \eAspectj{} was used to weave monitoring aspects (tracing) into the subject systems to capture their \emph{Runtime Behavior} in the modeling universe.
	\item The \emph{Probabilistic Model Network (empty)} was created by applying the rules from Section~\ref{sec: construction rules} for each code element. 
	Shape and size of the NVPs is given in Table \ref{tab:hyper parameters}.
	\item The \emph{Behavior Datasets} were created by tallying the event stream. 
	This includes splitting the dataset into training and evaluation partitions and preprocessing them.
	Preprocessing consisted of encoding text features by enumerating (starting from 0) and encoding them in a base 10 vector space.
	The same procedure was applied to the conditional dimension.
	Number dimensions were considered discrete if less or equal than \num{16} values were found and underwent the same base 10 encoding procedure.
	Finally, all dimensions were standardized to have a mean of zero and a standard deviation of 1.
	\item Model parameters were optimized with their datasets, and the best parameter setting was retained (w.r.t. evaluation performance).
	\item Finally, the persisted models were used in the analysis scenarios (see Section~\ref{sec:model analysis}).
\end{enumerate}

Hyper-parameters of the experiments are given in Table~\ref{tab:hyper parameters}.
The chosen values are based on additional non-reported experiments evaluated on a synthetic dataset.
All experiments were executed on a single machine (Intel i7, Nvidia GTX 970).

\subsection{Subject Systems}\label{sec: subject systems}
\input{tab_project_overview}

The study uses four subject systems listed in Table~\ref{tab:project overview}.
\emph{Nutrition Advisor} is the running example introduced in Section~\ref{sec:background}.
\emph{Structurizr}~\cite{Structurizr2019} is a developer-focused software architecture visualization tool.
\emph{jLatexmath}~\cite{Opencollab2019} is a library for rendering LaTeX formulas.
\emph{PMD}~\cite{PMD2019} is a static code analysis tool for Java applications.

All code elements of the projects were included in the modeling universe (excluding inherited third-party elements).
\emph{Nutrition Advisor} received \num{1000} advice requests as a trigger with data based on the NHANES \cite{Nhanes2013} dataset.
\emph{jLatexmath} and \emph{Structurizr} were executed with examples provided in their documentation.
\emph{PMD} analyzed the \emph{Nutrition Advisor} and output the results in HTML format.
The subject systems and their triggers are openly available\footnote{\ifunblind\url{https://github.com/jku-isse/gradient-benchmark}\fi} as a benchmark suite for future experiments and comparisons.

\subsection{Controlled Variables} \label{sec:controlled variables}
The study controls for one variable: \textit{Capacity}.
\begin{itemize}
	\item \textbf{Capacity:}\label{ep: capacity} The capacity describes the number ($\mathit{low}=32$, $\mathit{high}=128$) of units in the linear layers of the NVPs. 
\end{itemize}

\subsection{Response Variables}\label{sec:response variables}
The response is split into a quantitative and qualitative part.
The quantitative part evaluates the \textit{Events per Code Element (ECE)}, \textit{Distinct Values per Code Element (DCE)}, and \textit{Negative Log-Likelihood} (NLL).
The qualitative part assesses the visual fidelity of the samples generated by the model compared to the original dataset and evaluates the usefulness of the PSM network via a scenario-based evaluation given in Section~\ref{sec: inference analysis}.
\begin{itemize}
	\item \textbf{Events per Code Element (ECE):} \label{ep: active code elements} 
	Measures the number of events emitted by code elements.
	This provides insight into the runtime activity of elements and how many models need to be fitted.
	We report ECE1 and ECE10 to distinguish between dependencies/constants and real behavior carrying code elements.
	ECE1 includes all code elements with at least one event (all active code elements at runtime).
	ECE10 includes only code elements that emitted at least 10 events at runtime.
	\item \textbf{Distinct Values per Code Element (DCE):} \label{ep: distinct values}
	Measures the number of distinct values emitted by code elements.
	This provides insight into the capacity models must have.
	We report DCE1 and DCE10 where DCE10 includes code elements with at least 10 distinct values.
	\item \textbf{Average Negative Log-Likelihood (NLL):} \label{ep: loss} Measures the average Negative Log-Likelihood (Equation \ref{eq: nvp logp}) of data points under the model in natural units of information (nats; lower is better).
\end{itemize}

\subsection{Experiment Results}\label{sec:experiment results}
The study results are split into four groups: \emph{Code}, \emph{Runtime}, \emph{Modeling}, and \emph{Inference}.

\subsubsection{Code} \label{sec:code analysis} 
The projects contained a total of \num{27804} property, parameter, and executable code elements.
PMD is the largest project containing \SI{76}{\percent} of the total code elements. Nutrition Advisor is the smallest project containing \SI{0.25}{\percent}.
Most elements were executables (\SI{43}{\percent}) or parameters (\SI{39}{\percent}).
\SI{42}{\percent} of the elements were \emph{data} elements, i.e., had either a number or text type that is eligible for PSM modeling.
\SI{22}{\percent} were references within the modeling universe and the remaining \SI{36}{\percent} were elements of unknown type that were not within the modeling universe.
Table~\ref{tab:project overview} shows detailed results per subject system, element type, and data type.

\subsubsection{Runtime}\label{sec:monitoring analysis}
\input{tab_monitoring.tex}
Monitoring sessions lasted for a median duration of \SI{136.55}{\second} (\IQR{3.27}{369.35}) and were concurrently executed with the modeling sessions of other projects.
The median processing speed was \num{25101} events per second (\IQR{24727}{26283}).

During the monitoring session, a total of \num{24868732} events were emitted from \num{6002} code elements (\SI{22}{\percent} of total code elements).
\SI{36}{\percent} of the \num{6002} code elements emitted \emph{data} (text or number) events.
\SI{68}{\percent} were generated by the PMD project, while the least events were generated by the Nutrition Advisor \SI{0.12}{\percent}.
\SI{87}{\percent} of the events were \emph{data} (text or number) events while the remaining \SI{13}{\percent} were either \emph{reference} or \emph{unknown} events.

The \emph{event analysis} shows that most of the events (\num{24861389}) occurred on \num{3868} (\SI{14}{\percent} of total) code elements.
This excludes elements that emitted less than 10 events (ECE10).
\SI{36}{\percent} of the \num{3868} code elements generated \emph{data} (text or number) events.
Percentages for the largest and smallest, as for the data types match those of the events.
Differences are given in Table \ref{tab: monitoring} in terms of the central tendencies.

The \emph{distinct value analysis} shows that a total of \num{176052} distinct values were generated by \num{914} code elements (\SI{3.29}{\percent}).
This excludes elements that emitted less than 10 events (DCE10).
\SI{44}{\percent} of the \num{914} code elements generated \emph{data} events.
Most of the distinct values come from the PMD project that make up \SI{53}{\percent}.
Least distinct values were generated by the Structurizr with \SI{3.32}{\percent}.
Distinct values related to \emph{Data} were encountered  \SI{34}{\percent} while others were encountered \SI{66}{\percent} of the time.

\subsubsection{Modeling}\label{sec:model analysis}
\input{tab_model_overview}

Table \ref{tab: model overview} contains the detailed results of the low capacity setting and the margins for the high capacity setting.
The total wall time to optimize the parameters of all models was \SI{195}{\minute} (\SI{111}{\minute} for high capacity).
The median time one model needed to optimize in the low capacity setting was \average{72.42}{55.21}{93.16} (\average{38.60}{29.11}{50.16} for high capacity).

A total of \num{774} models were fitted. PMD accounted for \SI{74}{\percent} of the models.
In sum, \num{680080} data points were used in the process were Nutrition Advisor had the most data points available per model (\num{1000}).
A total of \num{3480} dimensions exist across all models were PMD accounts for \SI{72}{\percent} of all dimensions.
However, the Nutrition Advisor models had the highest amount of dimensions per model.
\SI{62}{\percent} of the dimensions were related to continuous features and the remainder to discrete features.
A total of \num{12787800} parameters were used (\average{15780}{15000}{16560}) in the low capacity setting for the models.
The high capacity setting had a total of \num{165172056} parameters (\average{210468}{207384}{213552}).
Finally, all projects yielded a total test NLL of \num{-3677.88} (low capacity).
On average, the models found in the PMD project had the best NLL with \num{-3.96} and the worst in the Structurizr \num{-0.93} (lower is better).
No significant divergence between training and test NLL can be seen.

The qualitative inspection of the models revealed a good approximation with two caveats.
First, imprecisions in the approximations are given for categorical dimensions that include high mass levels.
The high mass levels cause an increase of mass in the surrounding levels compared to the original data.
Proximity in categorical data is introduced by the 10-ary encoding and the continuous nature of NVPs.
Second, imprecisions are given in continuous dimensions with disconnected high-density modes being connected.
This issue occurs more frequently in the low capacity setting than in the high capacity setting indicating underfitted models.

\subsubsection{Inference} \label{sec: inference analysis}
\begin{figure}
	\centering
	\includegraphics[width=\linewidth]{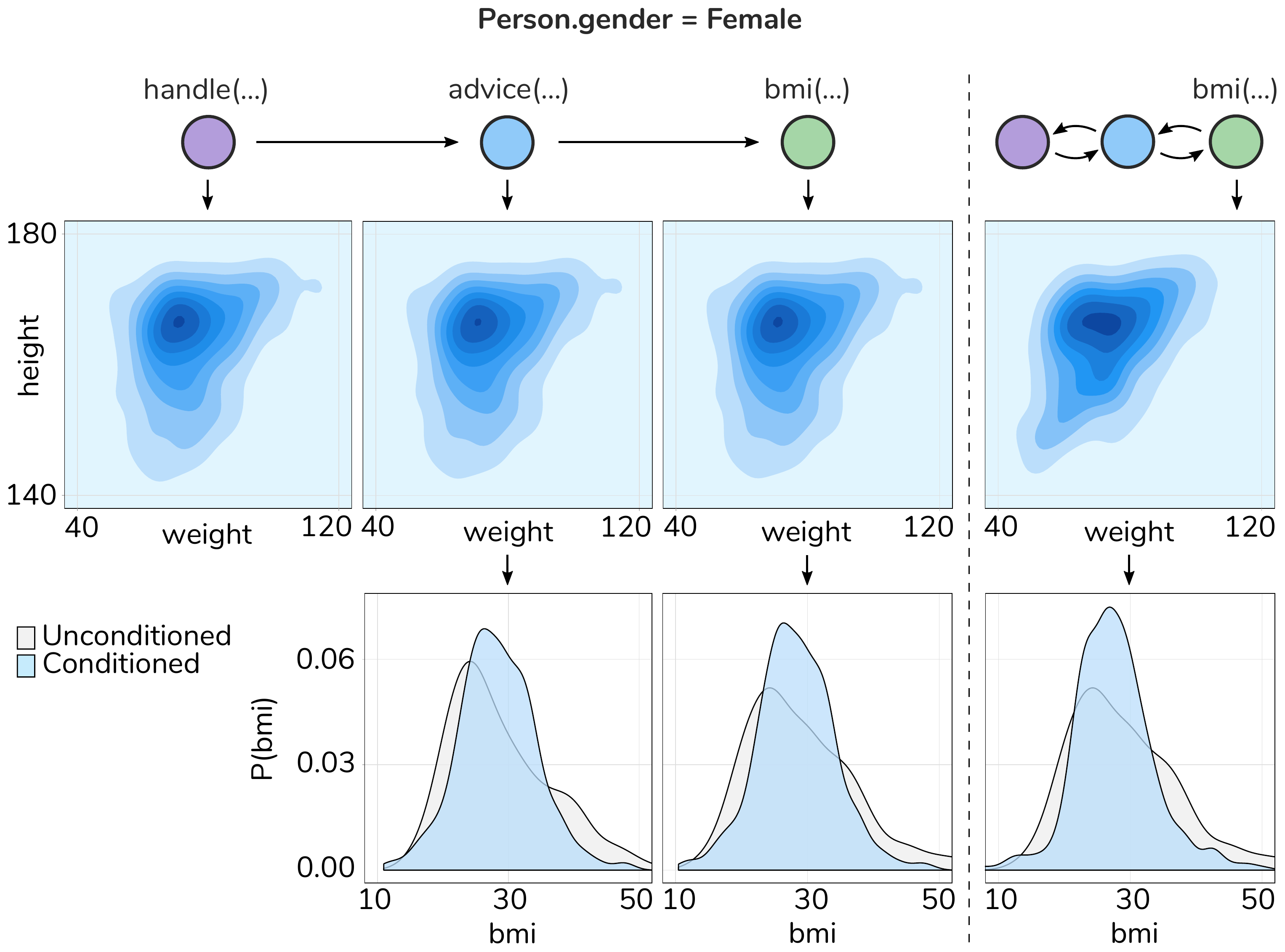}
	\caption{Shows an inference example with a condition caused by a latent variable starting at the handle-method.
	Gender, only accessible in the handle-method is conditioned to females.
	Height and weight are propagated while bmi jointly adapts to the condition. 
	The last column shows a roundtrip of 10 (40 propagation hops) and its effect on compared to the original distribution.}
	\label{fig: forward conditioning}
\end{figure}

\begin{figure}
	\centering
	\includegraphics[width=.8\linewidth]{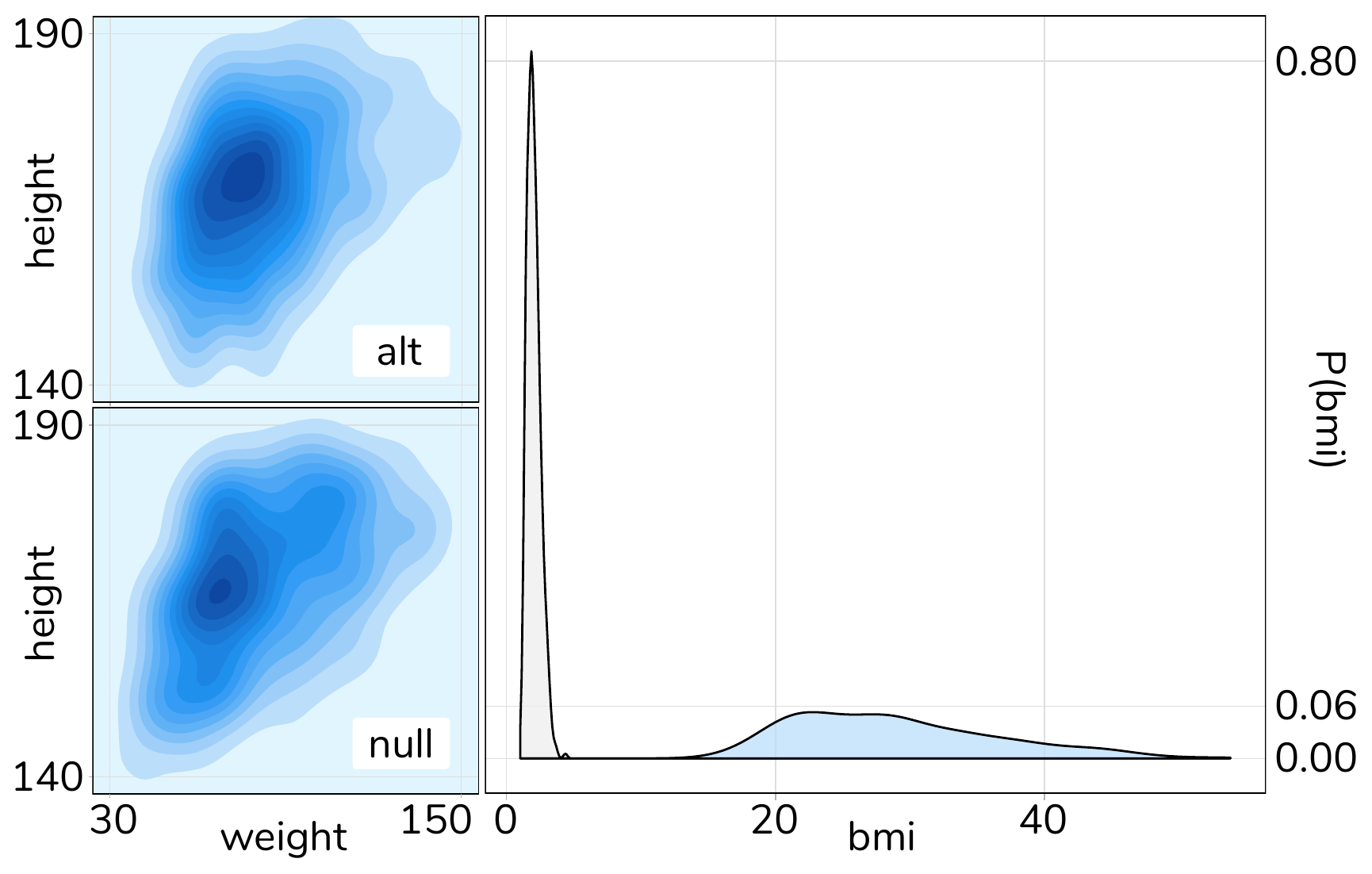}
	\caption{Shows an example for semantic testing and criticism where the \emph{null}-model and \emph{alt}-model come from different teams.
		The clear difference between the return values was detected automatically and works indifferent with traditional software as with software 2.0.
	}
	\label{fig: semantic equality}
\end{figure}

The qualitative assessment of the inference capabilities of PSM are split into two scenarios presented in Figure \ref{fig: forward conditioning} and Figure \ref{fig: semantic equality}.
These scenarios extend the running example by adding the \code{Servlet} to the Modeling Universe.

The first scenario in Figure~\ref{fig: forward conditioning} shows a simulation in which the Nutrition Advisor is conditioned on women requests.
The circles at the top illustrate the original call hierarchy and parts of the PSM network from Figure~\ref{fig: nutrition advisor structural}.
Each node was fitted on the original data without any restrictions or conditions.
The contour plots below show the \emph{height} and \emph{weight} variables in each model conditioned by gender (see Figure~\ref{fig: semantic equality} for unconditional version).
The density plots at the bottom present the \emph{bmi} variable of the same respective model.
In the background is the original unconditioned distribution (i.e., including males).
\textbf{Only} the \emph{handle}-model has direct access to the \emph{gender} property.
By iteratively sampling $n$ observations, propagating, and conditioning the next model the original conditional information (i.e., $Person.gender = Female$) flows through the network.
This equals $n$ (probabilistic) executions of the program.
Finally, Figure~\ref{fig: forward conditioning} on the right shows the degree of information degradation in a forward and backward inference setting with 10 round-trips (40 information hops).
Centers and shape are mostly preserved but a slight shift of variance can be seen.
The density of the \emph{bmi} variable was preserved over the 40 hops without any crucial loss of information.

The second scenario in Figure~\ref{fig: semantic equality} assumes that \emph{Servlet} and \emph{NutritionAdvisor} are developed by Company A while \emph{BmiService} is developed by Company B specialized on AI.
Company A uses the simple height/weight formula to stub the \emph{BmiService} until Company B delivers its service based on a regression model.
Company A has a PSM model $M_{null}$ of the system.
Company A builds a second revision $M_{alt}$ of its PSM model, including the new component they received from Company B (BmiService).
The automated compatibility checks during continuous integration failed for \emph{bmi} code elements (in \emph{bmi(\ldots)} and \emph{advice(\ldots))} but are successful for all other elements.
Revisiting the call graph in reverse order reveals a semantic error in the new component illustrated in Figure~\ref{fig: semantic equality}.
The inputs match (contour plots on the left) but the outputs diverge drastically (density plot on the right).
The issue was that Company A uses the metric measurement system while Company B uses the imperial system.

The scenario is based on real data. 
However, the regression model was substituted by the simple BMI formula given in the imperial form.
Compatibility checks were done with Kolmogorov-Smirnov Tests~\cite{Massey1951}.

The remarkable aspect of this scenario is the ignorance of PSM regarding the true underlying implementation (code vs AI model).
Unit tests of the component and integration tests of depending components would need to ask the model for the correct assertion values given an input.
Not only are these tests flawed, but every update of the model's parameters would trigger cascading changes in the tests.
In contrast, PSM tests the behavior, not the code (semantic tests).

\section{Discussion}\label{sec:discussion}
The results presented in Section~\ref{sec:experiment results} provide direct or indirect evidence for the research questions in Section~\ref{sec:study}.

\subsection{Code}
The results of the code analysis (see Section~\ref{sec:code analysis}) shows that the total project size is secondary for PSM.
Nearly half (\SI{43}{\percent}) of the code elements in a project are text or numbers and can be modeled.
The remaining elements are either referencing eligible code elements or are external dependencies.
This large proportion justifies the use of PSM for projects independent of their size (\ref{rq:code}).

In conclusion, projects, independent of their size, expose enough code elements eligible for PSM.

\subsection{Runtime}
The results of the runtime analysis (see Section~\ref{sec:monitoring analysis}) show that most events are related to actual data (\SI{87}{\percent}), providing evidence for \ref{rq:monitoring} and support for PSM.
These data events are emitted by a rather small portion of the active code elements (\SI{14}{\percent}, ACT10).
Regarding \ref{rq:modeling}, this means that few models will capture most of a program's behavior.
Most of the variability is generated by few code elements \SI{3.29}{\percent}.
Nearly half of the variability is related to data (\SI{44}{\percent}) while the other half are mostly object references.
In terms of \ref{rq:modeling} this means that the average capacity (free optimizable parameters) of models can be low; simplifying model maintenance and interpretation.

In conclusion, active code elements are creating enough data (text or number) that can be used for PSM.

\subsection{Modeling}
The results of the modeling analysis (see Section~\ref{sec:model analysis}) show that most models have few dimensions providing further empirical support to use low capacity models.
The selected capacity does not hint at overfitting to specific portions of the data given that training and test NLL are not significantly different.
However, many low-dimension discrete only models can be replaced by Conditional Probability Tables (CPDs)\footnote{A table encoding the probability per categorical level.}\cite{Koller2009} for a more efficient and precise representation.

The qualitative inspections revealed high-quality models with good approximations with two caveats (mass leakage and mode connectivity).
The two issues are related to the capacity of the model (too high for discrete, too low for continuous) that adaptive model type and parameter selection can solve.

In conclusion, the qualitative and quantitative assessments suggest that probabilistic models can approximate the behavior of a program.

\subsection{Inference}
The inference analysis (see Section~\ref{sec: inference analysis}) evaluated the usefulness of PSM models by two illustrative scenarios.
The first scenario (Figure~\ref{fig: forward conditioning}) illustrated multi-dimensional information (height and weight) propagation with latent factors (gender only visible in \code{request}) across multiple models.
The second scenario (Figure~\ref{fig: semantic equality}) focused on model/data evaluation in a software development context in which software and AI components are integrated.
The scenarios distill the foundations on which any PSM application (see Section~\ref{sec:applications}) is built: sampling (generation), conditioning (information propagation), and likelihood evaluation (criticism).

In conclusion, results show that local (within model) and global (between models) generation is sensitive to conditions allowing consistent causal reasoning in PSM models.

\section{Limitations}
There are several limitations to the approach or the current prototype.
The approach needs a structured program, and it must be observable at runtime.
Large methods that handle multiple tasks will reduce the usefulness of PSM.

The current prototype is focused on data.
References are handles to objects that might contain data or more references.
PSM naturally dereferences these handles since models only contain, e.g., properties, that are accessed.
This means that PSM is not useful for libraries whose only purpose is reference management, e.g., a collection library.

The current prototype explodes lists as singular value assignments, i.e., a list of two elements acts as two assignments to a non-list variable.
No order relationship between list elements is preserved as typical for distributions.
Sequential models can alleviate this limitation.
However, the usefulness is subject to the actual application that is realized.

\section{Threats to Validity}
An external threat to validity is given by the number of projects used in the study.
Rigorous internal evaluation and projects of different size and type minimize the threat.
Different sizes control for the expectation that large projects will have more elements and events, resulting in better models.
Different project types (e.g., PMD as system or jLatexmath as application software) control for the element type distribution and their runtime content (user vs. synthetic data).
Finally, the evaluation models \emph{all} eligible code elements and measured the variance across the projects.
The NLL across projects in Table~\ref{tab: model overview} does not hint at a by-chance good project selection.

\section{Conclusion and Future Work}\label{sec:conclusion}
In this work, we presented Probabilistic Software Modeling (PSM), a data-driven approach for predictive and generative methods in software engineering.

We have discussed applications, pragmatics, construction details, and technical considerations of PSM.
We evaluated the viability and usability of PSM on multiple projects and discussed scenarios that provide insight into how PSM is used.
The results have shown that PSM is not only viable but naturally integrates with software 2.0 (AI components).

Our future work will focus on the realization and evaluation of applications and their comparison to the current state-of-the-art.

In conclusion, PSM analyzes a program and synthesizes a probabilistic model that is capable of simulating and quantifying it.
The resulting models are repeatable, persistable, shareable, and quantifiable representations and act as a foundation from which solutions can be derived.

%% file: tab_model_parameters.tex
\begin{table}[!t]
    \scriptsize
    \centering
    \ra{1.2}
    \caption[Hyper-Parameters]{Hyper-parameters used in the experiments.}
    \label{tab:hyper parameters}
    \begin{threeparttable}
        \begin{tabular}{@{}c l r r @{}}
            \toprule
            \textbf{\#} & \textbf{Stage} & \textbf{Name} & \textbf{Values}\\
            \midrule
            1 & Data & Size & \numrange{20}{10000} \\
            2 & Data & Test Split & \num{10}\% \\
            3 & Preprocessing & Number & Standardization \\
            4 & Preprocessing & Discretization Threshold & 16 \\
            5 & Preprocessing & Discretization Encoding & Base 10 \\
            6 & Preprocessing & Text Encoding & Base 10 \\
            7 & Optimizer & Algorithm & Adam \cite{Kingma2014} \\
            8 & Optimizer & Learning Rate & \num{5d-4} \\
            9 & Optimizer & Weight Decay \cite{Krogh1992} & \num{5d-2} \\
            10 & Optimizer & Batch Size & full dataset \\
            11 & Optimizer & Max Epoch & \num{1000} \\
            12 & Optimizer & Early Stopping Patience & 20 epochs \\
            13 & NVP \cite{Dinh2016} & Coupling Count & \num{6} \\
            14 & Coupling Layer \cite{Dinh2016} & Linear Layer Count & \num{2} \\
            15 & Coupling Layer \cite{Dinh2016} & Hidden Units Count & \num{32} (low) \num{128} (high) \\
            16 & Coupling Layer \cite{Dinh2016} & Latent-Space & $\mathcal{N}(0, \bm{1})$ \\
            17 & Coupling Layer \cite{Dinh2016} & Translation Activations  & Gelu \cite{Hendrycks2016} \\
            18 & Coupling Layer \cite{Dinh2016} & Scale Activations  & Gelu \cite{Hendrycks2016}, Tanh \\
            \midrule
            \bottomrule
        \end{tabular}
    \end{threeparttable}

\end{table}

%% file: tab_project_overview.tex
\begin{table*}[!t]
    \scriptsize
    \centering
    \ra{1.2}
    \caption[Project Overview]{
        Overview of the projects used in the study.
        LoC are the lines of code in a project.
    }
    \label{tab:project overview}
    \begin{threeparttable}
        \begin{tabular}{@{}l r r r r r r r r r r r r r r r r r r r r @{}}
            \toprule
            \textbf{Project} & \textbf{Version} & \textbf{\#Files} & \textbf{\#LoC} & \textbf{Type} & \phantom{a} & \multicolumn{4}{r}{\textbf{Property}}& \phantom{a}  & \multicolumn{4}{r}{\textbf{Parameter}} & \phantom{a} & \multicolumn{5}{r}{\textbf{Executable}} \\
            \cmidrule{7-10} \cmidrule{12-15} \cmidrule{17-21}
            & & & & & \phantom{a}                                               & Data & Ref & Unk & Total   & \phantom{a}   & Data & Ref  & Unk & Total & \phantom{a} & Data & Ref & Void & Unk & Total \\
            \midrule    
            Nutrition Advisor   & 0.1.0 & \num{5}    & \num{154}   & \num{5}    && \num{11}   & \num{3}   & \num{1}   & \num{15}                     && \num{19}   & \num{1}    & \num{0}    & \num{20}                && \num{10}   & \num{0}    & \num{19}   & \num{1}    & \num{30}    \\
            Structurizr         & 1.0.0 & \num{115}  & \num{9941}  & \num{123}  && \num{229}  & \num{85}  & \num{24}  & \num{338}                    && \num{725}  & \num{342}  & \num{26}   & \num{1093}              && \num{320}  & \num{302}  & \num{508}  & \num{20}   & \num{1150}  \\
            jLatexmath          & 1.0.7 & \num{156}  & \num{21369} & \num{191}  && \num{490}  & \num{121} & \num{81}  & \num{692}                    && \num{1115} & \num{556}  & \num{153}  & \num{1824}              && \num{269}  & \num{416}  & \num{511}  & \num{59}   & \num{1255}  \\
            PMD                 & 6.5.0 & \num{799}  & \num{89349} & \num{981}  && \num{1858} & \num{503} & \num{481} & \num{2842}                   && \num{2933} & \num{2910} & \num{1943} & \num{7786}              && \num{3222} & \num{719}  & \num{3445} & \num{2073} & \num{9459}  \\
            \midrule    
            &&                            \num{1075}  & \num{120813} & \num{1300}  && \num{2588} & \num{712} & \num{587} & \num{3887}               && \num{4792} & \num{3809} & \num{2122} & \num{10723}             && \num{3821} & \num{1437} & \num{4483} & \num{2153} & \num{11894} \\
            \midrule
            \bottomrule
        \end{tabular}
    \begin{tablenotes}
        \item Data = \{Number, Text\}, Ref = Reference, Unk = Unknown
    \end{tablenotes}
    \end{threeparttable}

\end{table*}

%% file: tab_monitoring.tex
\begin{table*}[!t]
    \scriptsize
    \centering
    \ra{1.2}
    \caption[Project Overview]{
    Events are the number of events observed at runtime. ACT10 are the number of events observed at runtime on code elements with at least 10 events.
    DCT10 are the number of distinct values on code elements with at least 10 distinct values.
    }
    \label{tab: monitoring}
    \begin{threeparttable}
        \begin{tabular}{@{}l r r r r r r r r r r r r r r r r @{}}
            \toprule
            \textbf{Project} & \textbf{Data Type} & \phantom{a} & \multicolumn{4}{r}{\textbf{Events}} & \phantom{a} & \multicolumn{4}{r}{\textbf{ACT10}} & \phantom{a} & \multicolumn{4}{r}{\textbf{DCT10}} \\
            \cmidrule{4-7} \cmidrule{9-12} \cmidrule{14-17}
            &  & \phantom{a}                                     & Mdn & Q1 & Q3 & Total & \phantom{a}  & Mdn & Q1 & Q3 & Total &  \phantom{a} & Mdn & Q1 & Q3 & Total \\
            \midrule            
            \multirow{2}{*}{Nutrition Advisor}      & Data      && \num{1000}  &  \num{1000}  & \num{1000} & \num{21000}       && \num{1000} & \num{1000} & \num{1000}  & \num{21000}         && \num{524}  & \num{363}  & \num{824}  & \num{8040}     \\
                                                    & Others    && \num{1000}  &  \num{252}   & \num{1001} & \num{9008}        && \num{1001} & \num{1000} & \num{1501}  & \num{9002}          && \num{1000} & \num{1000} & \num{1000} & \num{1000}     \\[+3pt]
            \multirow{2}{*}{Structurizr}            & Data      && \num{6}     &  \num{2}     & \num{17}   & \num{35852}       && \num{25}   & \num{16}   & \num{67}    & \num{35041}         && \num{21}   & \num{13}   & \num{46}   & \num{2514}     \\
                                                    & Others    && \num{12}    &  \num{3}     & \num{36}   & \num{58489}       && \num{34}   & \num{17}   & \num{104}   & \num{57607}         && \num{29}   & \num{16}   & \num{59}   & \num{3331}     \\[+3pt]
            \multirow{2}{*}{jLatexmath}             & Data      && \num{130}   &  \num{15}    & \num{526}  & \num{6415336}     && \num{274}  & \num{61}   & \num{1297}  & \num{6414919}       && \num{39}   & \num{18}   & \num{81}   & \num{24495}    \\
                                                    & Others    && \num{66}    &  \num{6}     & \num{530}  & \num{1377280}     && \num{257}  & \num{56}   & \num{1064}  & \num{1376553}       && \num{107}  & \num{30}   & \num{408}  & \num{42592}    \\[+3pt]
            \multirow{2}{*}{PMD}                    & Data      && \num{35}    &  \num{5}     & \num{154}  & \num{15069591}    && \num{117}  & \num{37}   & \num{267}   & \num{15068209}      && \num{39}   & \num{18}   & \num{91}   & \num{24511}    \\
                                                    & Others    && \num{18}    &  \num{5}     & \num{117}  & \num{1882176}     && \num{64}   & \num{20}   & \num{185}   & \num{1879058}       && \num{30}   & \num{16}   & \num{123}  & \num{69569}    \\
            \midrule
            & &                                                  & \num{21}    &  \num{5}     & \num{138}  & \num{24868732}    && \num{83}   & \num{25}   & \num{306}   & \num{24861389}      && \num{39}   & \num{17}   & \num{102}  & \num{176052}   \\
            \midrule
            \bottomrule
        \end{tabular}
        \begin{tablenotes}
            \item Mdn = Median, Q1/3 = Quartile
            \item Data = \{Number, Text\}, Others = \{Reference, Unknown\}
        \end{tablenotes}
    \end{threeparttable}

\end{table*}

%% file: tab_model_overview.tex
\begin{table*}[!t]
    \footnotesize
    \centering
    \ra{1.2}
    \caption{
    Model analysis results split across projects, and capacity.
    Lower is better for NLL results.
    }
    \label{tab: model overview}
    \resizebox{\textwidth}{!}{
    \begin{threeparttable}
        \begin{tabular}{@{}l l r r r r r r r r r r r r r r r r r r r r r@{}}
            \toprule
            \textbf{Capacity} & \textbf{Project} & \textbf{Models} & \phantom{a} & \multicolumn{4}{r}{\textbf{Data Points}} & \phantom{a} & \multicolumn{4}{r}{\textbf{Dimensions}} & \phantom{a} & \multicolumn{4}{r}{\textbf{Training NLL}} & \phantom{a} & \multicolumn{4}{r}{\textbf{Test NLL}} \\
            \cmidrule{5-8} \cmidrule{10-13} \cmidrule{15-18} \cmidrule{20-23} 
            & & & \phantom{a}                 & Mdn & Q1 & Q3 & Total & \phantom{a} &  Mdn & Q1 & Q3 & Total                                                      & \phantom{a} &  Mdn & Q1 & Q3 & Total            & \phantom{a}  & Mdn & Q1 & Q3 & Total\\
            \midrule
            \multirow{4}{*}{Low} & Nutrition Advisor      & \num{4}     && \num{1000} & \num{1000} & \num{1000}  & \num{4000}           && \num{6}  & \num{5} & \num{8} & \num{27}        && \num{-1.37}  & \num{-4.40}  & \num{1.92}   & \num{-4.44}      && \num{-1.61} & \num{-4.51}  & \num{1.69}   & \num{-4.80}     \\
            &Structurizr            & \num{50}    && \num{67}   & \num{31}   & \num{137}   & \num{14715}          && \num{3}  & \num{3} & \num{4} & \num{179}       && \num{-0.83}  & \num{-2.77}  & \num{1.75}   & \num{-48.86}     && \num{-0.93} & \num{-2.95}  & \num{2.08}   & \num{-39.27}    \\
            &jLatexmath             & \num{146}   && \num{393}  & \num{82}   & \num{1248}  & \num{206820}         && \num{4}  & \num{3} & \num{7} & \num{763}       && \num{-3.10}  & \num{-7.64}  & \num{1.06}   & \num{-617.12}    && \num{-3.10} & \num{-7.91}  & \num{1.29}   & \num{-598.81}    \\
            &PMD                    & \num{574}   && \num{133}  & \num{56}   & \num{337}   & \num{454545}         && \num{4}  & \num{3} & \num{5} & \num{2511}      && \num{-3.96}  & \num{-6.84}  & \num{-3.15}  & \num{-3080.96}   && \num{-3.96} & \num{-6.69}  & \num{-2.94}  & \num{-3034.99}   \\
            \midrule
            Low &                   & \num{774}   && \num{151}  &  \num{56}  & \num{472}   & \num{680080}         && \num{4}  & \num{3} & \num{5} & \num{3480}      && \num{-3.95}  & \num{-6.67}  & \num{-1.96}  & \num{-3751.38}   && \num{-3.95} & \num{-6.58}  & \num{-1.96}  & \num{-3677.88}   \\
            High &                  &    &&   &    &    &          &&   &  &  &                                   && \num{-3.95}  & \num{-7.22}  & \num{-2.03}  & \num{-3985.55}   && \num{-3.99} & \num{-7.30}  & \num{-1.99}  & \num{-3946.18}   \\
            \midrule
            \bottomrule
        \end{tabular}
    \end{threeparttable}
   }
    \begin{tablenotes}
        \item Mdn = Median, Q1/3 = Quartile
    \end{tablenotes}
\end{table*}